\documentclass[twocolumn,aps,prl,showpacs,superscriptaddress]{revtex4-1}
\usepackage{amssymb,amsmath}
\usepackage{graphicx}
\usepackage{color}
\usepackage[pdftex]{epsfig}
\usepackage{epstopdf}

\newcommand{\snc}[1]{\textcolor{black}{#1}}
\newcommand{\mae}[1]{\textcolor{black}{#1}}

\begin{document}

\def\simlt{\mathrel{\lower .3ex \rlap{$\sim$}\raise .5ex \hbox{$<$}}}
\def\simgt{\mathrel{\lower .3ex \rlap{$\sim$}\raise .5ex \hbox{$>$}}}

\title{Coherent quantum oscillations of a Si charge qubit}

\author{Zhan Shi}
\author{C. B. Simmons}
\author{Daniel. R. Ward}
\author{J. R. Prance}
\author{R. T. Mohr}
\author{Teck Seng Koh}
\author{John King Gamble}
\author{Xian. Wu}
\author{D. E. Savage}
\author{M. G. Lagally}
\author{Mark Friesen}
\author{S. N. Coppersmith}
\author{M. A. Eriksson}
\affiliation{University of Wisconsin-Madison, Madison, WI 53706}


\begin{abstract}

Fast quantum oscillations of a charge qubit  in a double quantum dot fabricated in a Si/SiGe heterostructure are demonstrated and characterized experimentally.  The measured inhomogeneous
dephasing time $T_2^*$ ranges from 127~ps to $\sim$2.1~ns; it
depends substantially on how the energy difference of the
the two qubit states varies with external voltages, 
consistent with a decoherence process that is dominated by \mae{detuning noise} (charge noise
that changes the asymmetry of the qubit's double-well potential). 
\mae{In the regime with the shortest $T_2^*$,} applying a charge-echo pulse sequence increases
the measured inhomogeneous decoherence time from 127~ps to 760~ps,
demonstrating that low-frequency noise processes are an important dephasing mechanism.
\end{abstract}

\maketitle

Fast, coherent control of charge qubits
has been demonstrated in both superconducting circuits~\cite{Nakamura:1999p786,Yamamoto:2003p941} and III-V semiconductor quantum dots~\cite{Hayashi:2003p226804,Petersson:2010p246804}.  Beyond its intrinsic interest, understanding semiconductor charge qubit coherence is also important for spin qubits: mixing spin with charge degrees of freedom (either through spin-orbit coupling~\cite{Nowack:2007p1430} or the exchange interaction~\cite{Loss:1998p120,Kane:1998p133,Vrijen:2000p012306,DiVincenzo:2000p1642,Shi:2012p140503,Koh:PRL2012}) enables faster spin manipulation than would otherwise be possible.  When this mechanism is used, charge coherence can determine the ultimate fidelity of a spin qubit~\cite{Koh:PRL2012}.

\begin{figure*}
\centering
\includegraphics[width=1\textwidth]
{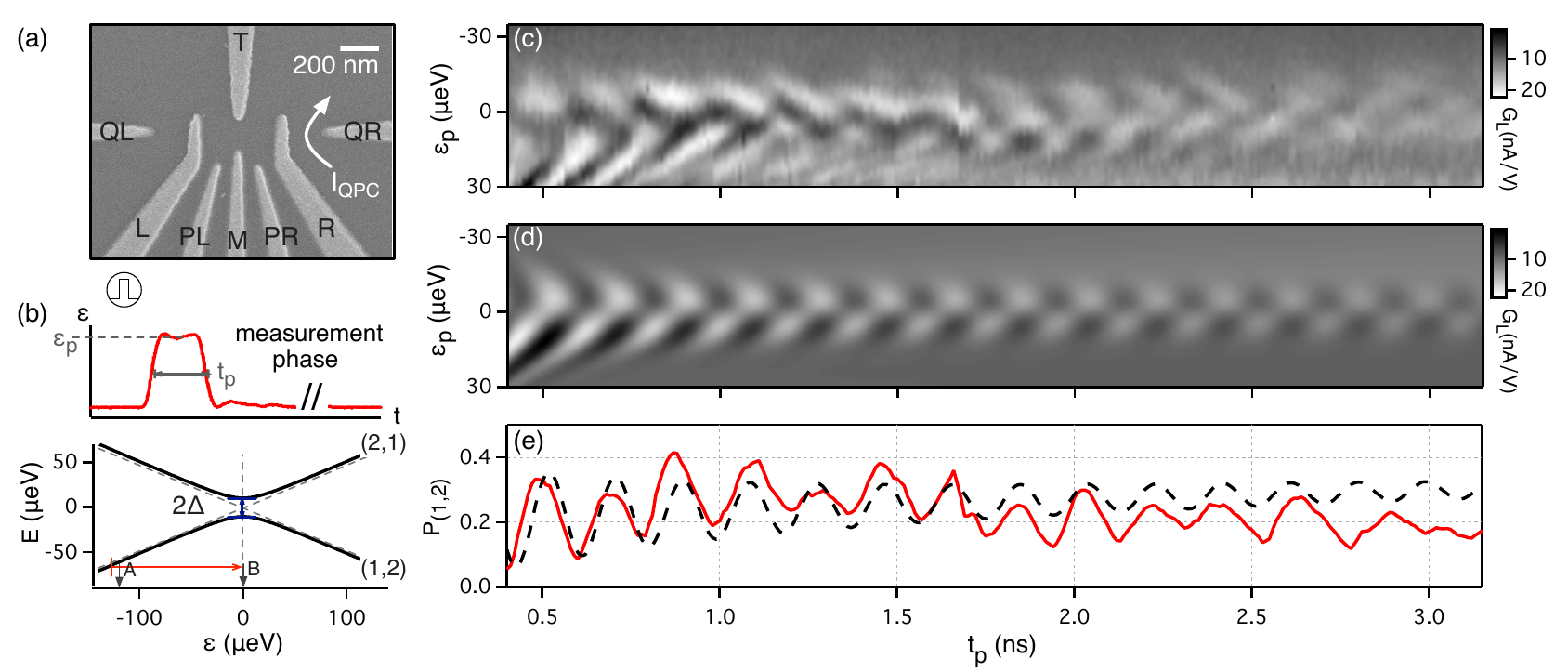}
\caption{\label{fig1}
\textcolor{black}{Larmor oscillations.}
(a) SEM image of a device identical to the one used in the
experiment. The current $I_\mathrm{QPC}$ is used for charge sensing through a measurement of its transconductance $G_\mathrm{L} = \partial I_\mathrm{QPC}/\partial V_\mathrm{L}$, and voltage pulses are applied to gate L.
(b) Bottom: Diagram of energy levels versus detuning $\varepsilon$, showing
the anticrossing between (2,1) and (1,2) charge configurations.
The
red arrow \textcolor{black}{represents} 
an applied voltage pulse with {the dc level and amplitude of the pulse} chosen to place the peak of the pulse at $\varepsilon = 0$.  Top: Example pulse showing the amplitude $\varepsilon_\mathrm{p} $, the pulse duration $t_\mathrm{p} $, and the measurement phase.  $t_\mathrm{p} $ varies from 400~ps to 3.15~ns, and the total pulse period, including the measurement phase, is 25~ns.
(c) Larmor oscillations between the (2,1) and (1,2) charge configurations, observed by 
performing a measurement of the QPC
transconductance $G_\mathrm{L}$ in the presence of a square pulse of duration $t_\mathrm{p}$ as a function of $t_\mathrm{p}$ and the position of the peak of the pulse on the detuning axis, $\varepsilon_\mathrm{p}$.  
The pulse amplitude is calibrated to be $V_\mathrm{p} = 5.27$~mV on gate L, and the pulse repetition rate is $40$~MHz.  
The Larmor oscillations that reflect rotations between the states with (2,1) and (1,2) charge occupations are
manifest near $\varepsilon_\mathrm{p} = 0$.
(d) Numerical simulation of the Larmor oscillations of \mae{(c)}, using \textcolor{black}{the measured} $80~\mathrm{ps}$ pulse rise time and the energy level diagram in (b) with 
best fit parameter $\Delta = 10.8$~$\mu$eV ($\Delta/h = 2.62$~GHz).
The calculation \textcolor{black}{includes explicit
high-frequency
dephasing parameterized by a dephasing time
$T_2^* = 2.1$~ns, and also incorporates low-frequency  fluctuations in the detuning $\varepsilon$ following Ref.~\cite{Petersson:2010p246804},} by convolving the calculated
evolution with a Gaussian of width $\sigma_\varepsilon = 5$~$\mu$eV, \textcolor{black}{as described in the main text}.
(e) Solid red: a horizontal cut taken near zero detuning from the integrated data of (c), showing Larmor oscillations as a function of pulse duration. Dashed black: a corresponding cut from the simulated charge occupation.
}
\end{figure*}

\begin{figure*}
\includegraphics[width=1\textwidth]
{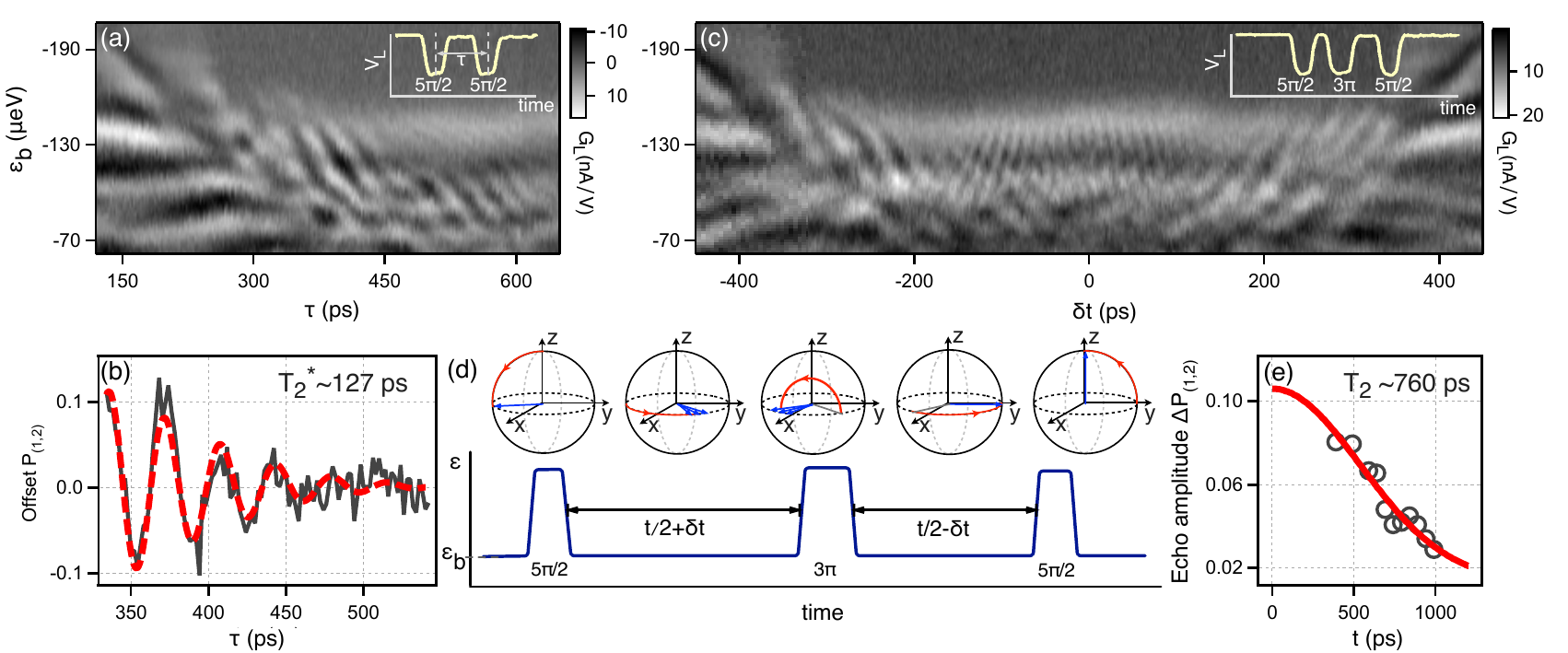}
\caption{\label{fig2}
(a) Ramsey fringes: QPC transconductance $G_\mathrm{L}$ as a function of the base level of detuning $\varepsilon_\mathrm{b}$ and the time $\tau$ between two $5\pi/2$ pulses, as shown in the inset.  The oscillations observed in the region where $\tau>280$~ps reflect the rotation of the Bloch vector around $z$ axis in the $x-y$~plane.
(b) Gray: a line cut of the integrated data after removal of a smooth background \cite{Supplemental:ZhanShi}.
Red: fit to the form
$A\exp (-(\tau-\tau_0)^2/T_{2}^{*2}) \cos (\omega t + \phi) + C$, which yields $T_2^* = 127 \pm 8$~ps.
(c) Charge echo: QPC transconductance $G_\mathrm{L}$ as a function of $\varepsilon_\mathrm{b}$ and $\delta t$ for $t=640$ps.  Oscillations are strongest near $\delta t=0$, where equal free evolution times before and after the $3\pi$ $x$-rotation provide the best correction for slow inhomogeneous dephasing.  Away from $\delta t = 0$, increasingly mismatched free evolution times provide less correction and the oscillations decay with characteristic time $T_2^*$. Inset: Trace of the pulse sequence used to acquire {these} data.
(d)  Sequence of pulses used to implement charge-echo: a nominal $5\pi/2$ $x$-rotation into the $x-y$ plane, free evolution for a time $t/2 + \delta t$, a $3\pi$ $x$-rotation, free evolution time $t/2 - \delta t$, and a $5\pi/2$ $x$-rotation out of the $x-y$ plane. For this experiment, $5\pi/2$ pulses have a duration of $280$~ps and $3\pi$ pulses have a duration of $330$~ps.
(e) Dark circles: charge oscillation amplitudes $\Delta P_{(1,2)}$as a function of the free evolution time $t$, extracted from data sets with different $t$ values. {A detailed description of the method used to obtain $\Delta P_{(1,2)}$ can be found} in~\cite{Supplemental:ZhanShi}.  The echo {amplitude decays as $t$ increases. A fit of the decay to the Gaussian form $y_0 + A\exp(-(t/T_2)^2)$ yields } $T_2 = 760 \pm 190$~ps, significantly longer than $T_2^*$.  {Thus,} the echo pulse sequence corrects for slow inhomogeneous {dephasing and extends the coherence time}.}
\end{figure*}

Here, we {present the first experimental measurements of} fast coherent quantum oscillations between the (2,1) and (1,2) charge states of a qubit formed in a Si/SiGe double quantum dot.
\snc{
Given a Bloch sphere~\cite{NielsenBook} with the $\pm z$ axes representing the (1,2) and (2,1) states,
rotations about the $x$-axis, or}
\snc{Larmor oscillations, are observed when the (2,1) and (1,2) states are energetically degenerate,
with a decoherence time $T_2^* = 2.1$~ns.
Rotations about the $z$-axis \mae{are probed in a Ramsey fringe experiment} and are observed with a shorter coherence time,}
\snc{$T_2^* = 127$~ps}.
The Ramsey fringes are measured when the charge qubit is operated {in} a {regime} {where}  the energy difference between the qubit
states depends strongly on detuning $\varepsilon$, \mae{the energy difference between the (2,1) and (1,2) charge states,} and so is highly sensitive to charge noise in the local environment.  The significantly different coherence times are consistent with the dominant dephasing mechanism arising from
fluctuations in $\varepsilon$, as has previously been observed in GaAs devices~\cite{Petersson:2010p246804}.
We also present the first measurements of charge echo in a semiconductor quantum dot charge qubit, with the echo sequence {yielding} an increase of the coherence time from 127 ps to 760 ps.
The results demonstrate control of a silicon charge qubit, and they show that charge echo
can be exploited to improve its coherence.

The device measured in the experiment
was fabricated in a Si/SiGe heterostructure as described in Refs.~\cite{thalakulamAPL10,simmonsPRL11};
a scanning electron microscope image of an identical device  is shown in Fig.~\ref{fig1}(a). 
The quantum point contact on the right side of the double dot is {used as a charge sensor}.
Using magnetospectroscopy measurements~\cite{Shi:2011p233108}, we confirm the valence charge occupation of the double dot is (2,1) or (1,2); either or both dots may contain a closed shell beneath the valence electrons, although if present such shells do not appear to play a role in the work we report.

Fig.~\ref{fig1}(b) shows an energy level diagram of the anticrossing between (2,1) and (1,2) as a function of detuning energy $\varepsilon$.  
\mae{Near the charge degeneracy point ($\varepsilon=0$) 
the system is well-described} by the Hamiltonian of a two-state
system:
\begin{equation}
H = \left( 
\begin{array} {cc}
\varepsilon/2 & \Delta \\
\Delta & -\varepsilon/2 \\
\end{array} \right).
\end{equation}
Coherent oscillations between the two charge states
can be observed when the detuning $\varepsilon$ is changed
abruptly.
For example, starting in a position eigenstate with charge occupation (2,1) at large negative $\varepsilon$, after increasing the detuning suddenly to $\varepsilon=0$,
as shown by the red horizontal arrow in Fig.~\ref{fig1}(b), the
system Hamiltonian becomes $H = \Delta  \sigma_x$, where $\sigma_x$ is the usual Pauli matrix.
Subsequently, the system oscillates between (2,1) and (1,2) at the Larmor angular frequency $2\Delta/\hbar$.
More generally, non-adiabatically increasing the detuning $\varepsilon$ to a value $\varepsilon^\prime$
is expected to induce oscillations at the angular frequency $\Omega_\mathrm{R}  = \sqrt{{\varepsilon^\prime}^2
+ 4\Delta^2}/\hbar$ about a tilted axis: as one moves away from the polarization line at which
$\varepsilon=0$, the oscillations
increase in frequency and decrease in amplitude.




Figure~\ref{fig1}(c) shows a number of Larmor oscillations between the (2,1) and (1,2) charge states.  Square pulses of duration $t_\mathrm{p}$ and amplitude $V_\mathrm{p} = 5.27$~mV on gate L are applied at frequency $40$~MHz.  The transconductance $G_\mathrm{L} = \partial I_\mathrm{QPC}/\partial V_\mathrm{L}$
is plotted as a function of $t_\mathrm{p}$ and $\varepsilon_\mathrm{p}$, the position in detuning of the peak of the pulse.  (The connection between $V_\mathrm{L}$, $t_\mathrm{p}$, $\varepsilon_\mathrm{p}$, and other details can be found in~\cite{Supplemental:ZhanShi}.)
Oscillations of the signal are apparent out to more than three nanoseconds.
To enable comparison to theory, and to obtain quantitative dephasing times from the experiment, we integrate the data presented in Fig.~\ref{fig1}(c) from top to bottom and extract the
probability $P_{(1,2)}$ of occupying the (1,2) charge state. Fig.~\ref{fig1}(e) presents the charge oscillation near zero detuning from the integrated data of (c)  as the solid red trace. By fitting the amplitude of the oscillations to exponential decays, we extract a dephasing time $T_2^* = 2.1 \pm 0.4$~ns near $\varepsilon_\mathrm{p}=0$, marked by label B in Fig.~\ref{fig1}(b). 
\snc{This dephasing time is relatively long, because at the anticrossing  the difference in energy between the eigenstates is insensitive to detuning fluctuations}.  

Numerical simulations of the experiment were performed based on the energy level diagram in Fig.~\ref{fig1}(b)
using the measured pulse rise time of 80 ps.  
We model the dynamical evolution of the density matrix $\rho$ of the system as a function of detuning $\varepsilon$ and pulse duration $t_p$ using a master equation~\cite{NielsenBook, Breuer2002}:
\begin{equation}
\dot{\rho} = -\frac{i}{\hbar} [ H , \rho ] + D,
\end{equation}
where $D$ is a phenomenological term that describes pure dephasing of the charge state under the assumption of Markovian dynamics. $D$ is given, in the $\{ |(2,1)\rangle, |(1,2)\rangle \}$ basis, by
\begin{eqnarray}
D &=& - \left( 
\begin{array} {ccc}
0 & \Gamma_{0}~\rho_{12} \\
\Gamma_{0}~ \rho_{21} & 0 \\
\end{array} \right),
\end{eqnarray}
\textcolor{black}{where} $\Gamma_{0}=0.48$~GHz is the dephasing rate ($1/T_2^*$) measured near zero detuning, which presumably is \textcolor{black}{dominated} by the effect of charge noise on tunnel coupling, since the qubit is \textcolor{black}{to} first order insensitive to detuning noise at the charge degeneracy point.
\mae{The (2,1) and (1,2) occupation probabilities are extracted at the end of the $t_\mathrm{p}$-pulse and, for the duration of the measurement phase, are allowed to relax exponentially to the ground state (2,1) occupation with a relaxation time $T_1$.  The simulated charge occupation is determined by averaging the charge state for the entire 25~ns pulse period.}
Low-frequency fluctuations in the detuning $\varepsilon$ are incorporated following Ref.~\onlinecite{Petersson:2010p246804}, by performing a convolution
of the results at each $\varepsilon$ with a Gaussian in $\varepsilon$ of width  $\sigma_\varepsilon = 5~\mu$eV.  The best fit to the data is found with a charge $T_1=18$~ns.

To enable comparison with the transconductance data in Fig.~1(b), we convert the simulated charge occupation to a QPC current using \textcolor{black}{the measured} QPC sensitivity of 18pA/electron and the measured cross talk between $V_L$ and the current obtained from the experiment(8.87~nA/V). We then differentiate the current with respect to $V_L$ using the \textcolor{black}{measured} left dot lever arm $\alpha_{L,\varepsilon}=-24~\mu$eV/mV, to produce simulated transconductance data.
\textcolor{black}{The results obtained using the tunneling amplitude $\Delta = 10.8$~$\mu$eV ($\Delta/h = 2.62$~GHz), shown in Fig.~\ref{fig1}(d), agree well with the data.}
\textcolor{black}{Fig.~\ref{fig1}(e) is a horizontal cut showing the measured and simulated charge occupation data near $\varepsilon=0$.  Again, good agreement between the data and the calculation is found.}


We now demonstrate coherent rotations of
the charge qubit about the $z$-axis of the Bloch sphere by performing
a Ramsey fringe experiment~\cite{Vion:2002p886,Dovzhenko:2011p161802}, using the
two-pulse sequence shown in the inset to Fig.~\ref{fig2}(a).
Starting at a negative detuning in the (2,1) state, the qubit is pulsed
to the (2,1)-(1,2) anticrossing, which causes the Bloch vector to rotate around the $x$-axis.
The duration of this first pulse is chosen (based on the data in Fig.~1) so that the Bloch vector is rotated around the $x$-axis by a nominal angle of $5\pi/2$, taking it from being
 along $z$ to being in
{the $x$-$y$ plane} (we use a $5\pi/2$ pulse of amplitude $3.95$~mV on gate L and duration $280$~ps
instead of a $\pi/2$ pulse, because of the difficulty of applying high-quality pulses {shorter than $100$~ps}).
After a variable {free evolution} time $\tau$ at the base level of the detuning $\varepsilon_\mathrm{b}$, during which {the Bloch vector rotates about the $z$-axis}, a second pulse is applied {to rotate} the
state about the $x$-axis on the Bloch sphere by another $5\pi/2$.
The charge measured at the end of this process oscillates as a function
of the time $\tau$ between the two 
pulses at a frequency determined by the difference in energy of the states involved at the base level of detuning.

Fig.~\ref{fig2}(a) shows the transconductance $G_\mathrm{L}$ of the charge sensor as a function
of the base level detuning $\varepsilon_\mathrm{b}$ and the time $\tau$, in the presence of the two-pulse pattern applied at a repetition rate of $25$~MHz.
For very short $\tau$, the $5\pi/2$ pulses overlap and one is essentially
performing a Larmor oscillation experiment.
At $\tau \simgt 280$~ps, the time interval between the end of the first $5\pi/2$ pulse and the start of the second
becomes nonzero, and the observed oscillations correspond to a Ramsey fringe measurement. 

To analyze these data quantitatively, we again integrate the transconductance and normalize by noting that the total charge transferred across
the polarization line is one electron.
Fig.~\ref{fig2}(b) shows a cut through the integrated data at the value
of detuning { $\varepsilon_\mathrm{b}=-120\mu$eV} marked by the arrow labeled A in Fig.~\ref{fig1}(b), after subtraction of a smooth background~\cite{Supplemental:ZhanShi}.
These Ramsey fringes oscillate at {$28$~GHz}, which agrees with the energy difference of the two charge states at that detuning value. We fit these oscillations to the product of a cosine function and a
Gaussian.\cite{Petersson:2010p246804}  This procedure yields $T_2^* = 127 \pm 8$~ps, much shorter than $T_2^*=2.1$~ns measured
in the Larmor experiment.  As is clear from Fig.~\ref{fig1}(b), at the large negative detuning at which the
oscillations are being generated, the energy levels diverge rapidly from each other as a function of $\varepsilon$, providing no protection from charge noise.  The dephasing time in this Ramsey fringe
experiment is nonetheless
{twice as long as} the value
of $60$~ps obtained by measuring Ramsey fringes for a GaAs 
charge qubit and using a similar fitting procedure to extract $T_2^*$~\cite{Dovzhenko:2011p161802}.  Calculations following the methods of Ref.~\cite{gamblePRB2012}
show that the charge dephasing rate in GaAs from polar optical phonons may be of order 1~GHz, whereas similar calculations for phonon-induced charge dephasing in Si yield values of order 0.5~MHz.  Thus, in both materials, and particularly in Si, improvements may be possible through a reduction of excess charge noise.

We now demonstrate that
 the effects of inhomogeneous { dephasing} can be ameliorated using a charge-echo method~\cite{Nakamura:2002p47901}.
 Charge-echo is implemented
 by applying the voltage
pulse sequence shown in Fig.~\ref{fig2}(d).  When the tips of the pulses reach the (2,1)-(1,2) anticrossing, the pulse sequence consists of a $5\pi/2$ pulse (which rotates the Bloch vector into the $x$-$y$ plane),
a free evolution at the base detuning for a time $t/2+\delta t$, a $3\pi$ pulse {(which flips the Bloch vector to its mirror image with respect to $x$-$z$ plane), a second free evolution at the base detuning for a time $t/2-\delta t$}, and a second $5\pi/2$ pulse (which rotates the Bloch vector { about the $x$-axis again).}
Fig.~\ref{fig2}(c) shows the transconductance measured as a function of the detuning $\varepsilon_\mathrm{b}$
and the time $\delta t$ for $t = 640$~ps.
{The oscillations are strongest at $\delta t = 0$, where the echo sequence best corrects for inhomogeneous dephasing.  As $|\delta t|$ increases, more time is spent performing an uncorrected $z$-rotation, and the oscillation amplitude decays with characteristic time $T_2^*$, just as in the Ramsey fringe experiment.  
As the total free evolution time $t$ increases, the oscillation amplitude will decay with characteristic time $T_2$.}   {To extract $T_2$, we perform the echo pulse sequence for multiple values of $t$.  We convert the the transconductance data to charge occupation data and extract the amplitude of the charge oscillation $\Delta P_{(1,2)}$~\cite{Supplemental:ZhanShi}.  Fig.~\ref{fig2}(e) shows the extracted value of $\Delta P_{(1,2)}$ for each data set, plotted as a function of $t$. The echo amplitude clearly decays  as $t$ is made longer.}
{A Gaussian fit of the decay yields $T_2=760\pm 190 $~ps.}
\snc{The significant increase in coherence time indicates that low-frequency noise processes
play an important role in limiting qubit coherence.}

In summary, we have observed coherent quantum charge oscillations in a qubit formed in a Si/SiGe double quantum dot.  Both Larmor oscillations and Ramsey fringes are observed.  The coherence time $T_2^*$ is $\sim$2.1~ns for Larmor oscillations at the charge degeneracy point.
Ramsey fringes that reflect quantum oscillations about the $z$-axis of the Bloch sphere are also observed.
Implementation of a charge-echo pulse sequence increases the decoherence time from $127$ ps to $760$ ps in the
regime in which the energy difference between the two qubit states depends substantially on detuning.

We thank Sankar Das Sarma and Jason Petta for useful discussions.  This research was supported in part by the U.S. Army Research Office
(W911NF-08-1-0482,
W911NF-12-1-0607), the NSF (DMR-0805045), and by the United States Department of Defense.  The views and conclusions contained in this document are those of the authors and should not be interpreted as representing the official policies, either expressly or implied, of the US Government.  This research utilized NSF-supported shared facilities at the University of Wisconsin-Madison.

\bibliography{from-zwan.bib,siliconqcsnc,new}

\end{document}


\def\simlt{\mathrel{\lower .3ex \rlap{$\sim$}\raise .5ex \hbox{$<$}}}
\def\simgt{\mathrel{\lower .3ex \rlap{$\sim$}\raise .5ex \hbox{$>$}}}

\title{Supplementary Information for ``Coherent quantum oscillations of a Si charge qubit''}

\author{Zhan Shi}
\author{C. B. Simmons}
\author{Daniel. R. Ward}
\author{J. R. Prance}\affiliation{University of Wisconsin-Madison, Madison, WI 53706}
\author{R. T. Mohr}\affiliation{University of Wisconsin-Madison, Madison, WI 53706}
\author{Teck Seng Koh}\affiliation{University of Wisconsin-Madison, Madison, WI 53706}
\author{John King Gamble}\affiliation{University of Wisconsin-Madison, Madison, WI 53706}
\author{Xian. Wu}\affiliation{University of Wisconsin-Madison, Madison, WI 53706}
\author{D. E. Savage}\affiliation{University of Wisconsin-Madison, Madison, WI 53706}
\author{M. G. Lagally}\affiliation{University of Wisconsin-Madison, Madison, WI 53706}
\author{Mark Friesen}\affiliation{University of Wisconsin-Madison, Madison, WI 53706}
\author{S. N. Coppersmith}\affiliation{University of Wisconsin-Madison, Madison, WI 53706}
\author{M. A. Eriksson}\affiliation{University of Wisconsin-Madison, Madison, WI 53706}


\begin{abstract}
We (1) provide details of the measurements and analysis of the Larmor oscillations and Ramsey fringes, and (2) provide details of the analysis of the charge echo experiment.
\end{abstract}

\maketitle

\section{Measurement Details and Line Cuts of Larmor and Ramsey Oscillations}

The data shown in Fig.~1(c) of the main text is acquired by sweeping the voltage $V_\mathrm{L}$ on gate L as a function of the pulse duration $t_\mathrm{p}$.  As $t_\mathrm{p}$ increases, \mae{because of the high-pass filter in the bias-tee for the high frequency line,} the time-averaged voltage on gate L changes, resulting in a linear change in the relationship between $V_\mathrm{L}$ and detuning $\varepsilon$.  Thus, to remove this offset, before converting to $\varepsilon$ we shift each vertical scanline by an amount $\delta V_\mathrm{L} = V_\mathrm{p} \times t_\mathrm{p} \times f_\mathrm{rep}$, where $V_\mathrm{p}$ is the pulse amplitude and $f_\mathrm{rep}$ is the pulse repetition frequency.  The vertical axis in Fig.~1(c) then is converted to $\varepsilon_\mathrm{p}$, the detuning value at the peak of the pulse, by noting that the main, slowest oscillation corresponds to $\varepsilon_\mathrm{p} = 0$, and by fitting the change in Larmor frequency as a function of $V_\mathrm{L}$ for small positive $\varepsilon_\mathrm{p}$.
For our experimental conditions, one Volt applied to the high frequency coaxial line connected to gate L results in a change in gate voltage of
6.6~mV on that gate.
The connection between $V_\mathrm{L}$ and detuning $\varepsilon$ is $\alpha_\mathrm{L,\varepsilon} =  -24~\mathrm{\mu eV/mV}$ for the Larmor oscillation data and $\alpha_\mathrm{L,\varepsilon} =  -36~\mathrm{\mu eV/mV}$ for the Ramsey fringe and charge echo data (these data sets were acquired at different gate voltage tunings of the dots). The lever arm is determined by fitting the Larmor oscillation frequency as a function of gate voltage $f = \sqrt{{\alpha_\mathrm{L,\varepsilon}V_L}^2
+ 4\Delta^2}/h$. \\

\setcounter{subfigure}{1}
\begin{figure}
\includegraphics[width=.44\textwidth]{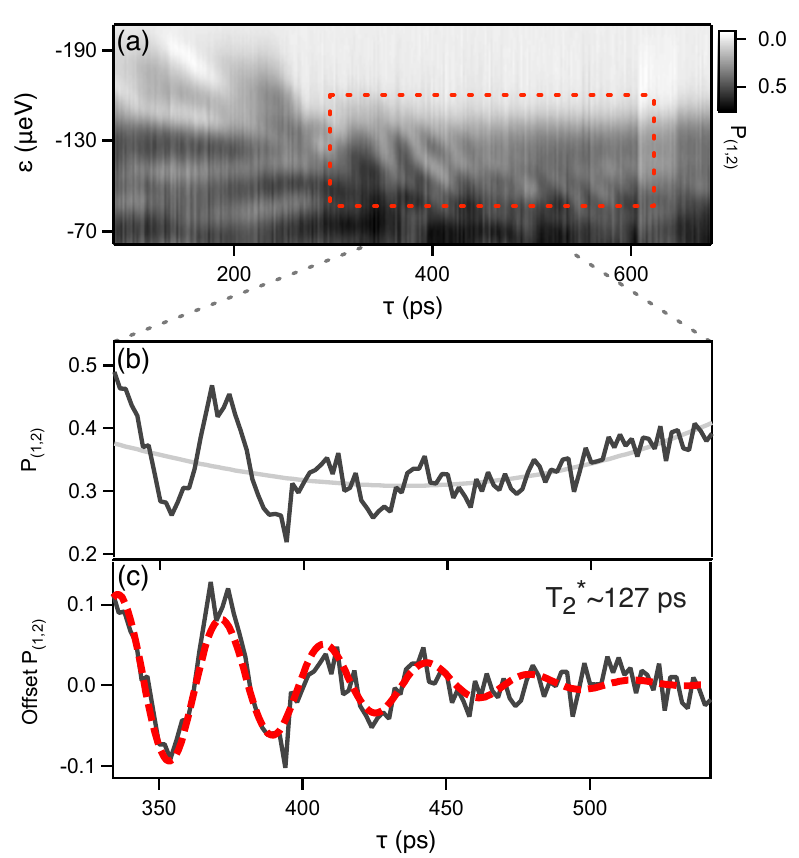}
\caption{\label{fig:plots4}
Ramsey fringe analysis.
(a) Integration of the transconductance $G_\mathrm{L}$ from Fig.~2(a) of the main text yields the probability $P_\mathrm{(1,2)}$ of occupying the (1,2) charge state in the regime where the (2,1) charge state is the ground state. The data is normalized by noting that the total charge transferred across the polarization line is one electron. The red dashed box indicates the location of the fringes. 
(b) Dark gray: line cut of the data in (a), at the detuning $\varepsilon=-120~\mu$eV.  Light gray: Smooth background that is subtracted from the line cut before fitting the data to a damped sinusoidal form. 
(c) (The same as Fig.~2(b) in the main text.)  Gray: the data from (b) after subtraction of the smooth background.
Red: fit to the form
$A\exp (-(\tau-\tau_0)^2/T_{2}^{*2}) \cos (\omega t + \phi) + C$, which yields $T_2^* = 127 \pm 8$~ps
 .}
\end{figure}

\setcounter{subfigure}{2}
\begin{figure*}
\includegraphics[width=0.85\textwidth]{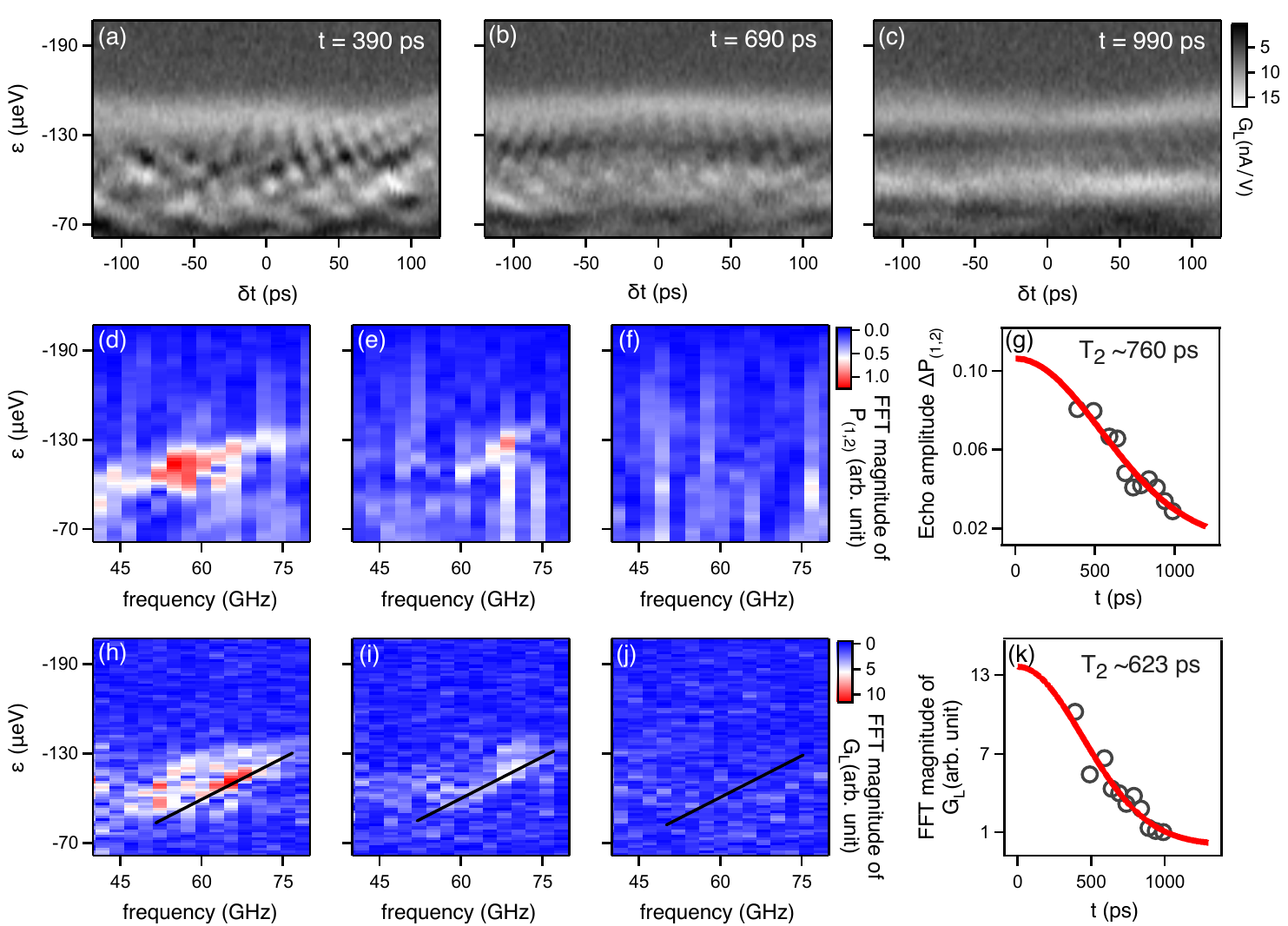}
\caption{\label{fig:plots5}
Analysis of echo data for extraction of the decoherence time $T_2$.
(a--c)  Transconductance $G_L$ as a function of the base level of detuning $\varepsilon$ and $\delta t$ (defined in the main text) for total free evolution times of $t=390$~ps, $t=690$~ps, and $t=990$~ps, respectively.  (d--f) Fourier transforms of the charge occupation $P_{(1,2)}$ as a function of detuning $\varepsilon$ and oscillation frequency $f$ for the data in (a--c), respectively.  We obtain $P_{(1,2)}$ (not shown here) by integrating the transconductance data in (a--c) and normalizing by noting that the total charge transferred across the polarization line is one electron.  Fast Fourier transforming the time-domain data of $P_{(1,2)}$ allows us to quantify the amplitude of the oscillations visible near $\delta t = 0$. The oscillations of interest appear as weight in the FFT that moves to higher frequency at more negative detuning (farther from the anti-crossing). For an individual detuning energy, the FFT has nonzero weight for a nonzero bandwidth. (g) Echo amplitude as a function of free evolution time $t$. The data points (dark circles) are obtained at $\varepsilon=-120$~$\mu$eV by integrating a horizontal line cut of the FFT data over a bandwidth range 
of $46$-$72$~GHz, then normalizing by the echo oscillation amplitude of the first data point, as described in the supplemental text. The echo oscillation amplitudes, plotted for multiple free evolution times, decay with characteristic time $T_2$ as the free evolution time $t$ is made longer. By fitting the decay to a Gaussian, we obtain $T_2 = 760 \pm 190$~ps.
(h--j) Fourier transforms of the transconductance $G_L$ as a function of $\varepsilon$ and oscillation frequency $f$ for (a--c), respectively.   As $t$ is increased, the magnitude for oscillations at a given frequency decays with characteristic time $T_2$.  We take the magnitude of the FFT at the point where the central feature (black line) intersects $65$~GHz.  (k) Measured FFT magnitudes at 65~GHz for multiple free evolution times (dark circles) with a Gaussian fit (red line), which yields $T_2 = 620 \pm 140$~ps, in reasonable agreement with the result shown in (g).}

\end{figure*}

Figure~\ref{fig:plots4}(a) shows the results of the integration of the data in Fig.~2(a) in the main text, normalized to obtain the probability $P_\mathrm{(1,2)}$ of being in the (1,2) charge configuration when (2,1) is the ground state.  Fig.~\ref{fig:plots4}(b) shows a line cut through the plot in Fig.~\ref{fig:plots4}(a) at the value of $\varepsilon=-120~\mu$eV.  Fig.~\ref{fig:plots4}(c), which is the same as Fig.~2(c) in the main text, shows the Ramsey fringes after subtraction of the smooth background shown in Fig.~\ref{fig:plots4}(b).

\section{Analysis of charge echo experiment}

To extract $T_2$ from the charge echo data we perform two analyses.  In the first, the oscillation amplitude is quantified at a given value of the detuning by analyzing the fast Fourier transform (FFT) of the probability $P_{(1,2)}$.  In the second, the oscillation amplitude is quantified at a given oscillation frequency by analyzing the FFT of $G_\mathrm{L}$.  As shown below, the results from the two methods are consistent.

The results presented in Fig.~2 of the main text were obtained by analyzing the oscillation amplitude of $P_{(1,2)}$ at fixed detuning.  To get $P_{(1,2)}$ as a function of detuning and $\delta t$, we integrate the time-domain data (such as that shown in Fig.~S2(a--c)) from top to bottom.  After removing a linear background, we normalize by noting that the total charge transferred across the polarization line is one electron.  We perform an FFT (using Igor Pro~\cite{wavemetrics}), and to ensure that the FFT magnitude is comparable for different values of $t$, we use the same number of points (or equivalently, the same length of time) from each data set by taking a $364$~ps cut centered about $\delta t \approx 0$.  The transforms are shown in Fig.~S2(d--f).  The oscillations of interest appear as spectral weight that moves to higher frequency at more negative detuning (farther from the anti-crossing). Moreover, for each value of the detuning, the FFT magnitude is nonzero over a certain range of frequencies.  

To extract the charge oscillation amplitude from these FFTs, we first take a horizontal trace from the FFT data at $\varepsilon=-120~\mu$eV, where the pulse tip is around zero detuning. We then integrate the trace over a bandwidth region from $46$ to $72$~GHz.  Because for the shortest free evolution time $t$ the oscillations in $P_{(1,2)}$ can be extracted easily from the untransformed data, we use that oscillation amplitude to normalize each of the FFT integrations, allowing us to plot a normalized oscillation amplitude as a function of $t$ in Fig.~S2(g).  The echo amplitude decays as the free evolution time $t$ is made longer, with a characteristic time $T_2$. Fitting a Gaussian to the decay yields $T_2 = 0.76 \pm 0.19$~ns.

For comparison, we also extract $T_2$ from an analysis of the FFT of the unintegrated transconductance data at fixed oscillation frequency.  We perform an FFT with a square window function on a $386$~ps cut centered at $\delta t \approx 0$ and plot the magnitude as a function of detuning and frequency, as shown in Fig.~S2(h--j).  We take the magnitude at the point where the central feature (black line) intersects 65 GHz and plot this quantity as a function of $t$ (Fig.~S2(k)).  Fitting to a Gaussian decay yields $T_2 = 0.62 \pm 0.14$~ns.  As Fig.~S2 demonstrates, the two approaches of extracting $T_2$ yield similar results.

\bibliography{from-zwan.bib,siliconqcsnc,new}